\documentclass[useAMS,usenatbib]{mn2e}
\usepackage{epsfig}
\usepackage{times}

\def\gtsim{\mathrel{\spose{\lower.5ex \hbox{$\mathchar"218$}}
     \raise.4ex\hbox{$\mathchar"13E$}}}
\def\ltsim{\mathrel{\spose{\lower.5ex\hbox{$\mathchar"218$}}
     \raise.4ex\hbox{$\mathchar"13C$}}}
\def\aFe{$\alpha/{\rm Fe}$}
\def\Hb{${\rm H}\beta$}
\def\Mgb{{\rm Mg}\, $b$}
\def\Fe{$\langle {\rm Fe}\rangle$}

\def\ZH{$Z/{\rm H}$}
\def\MgFe{$[{\rm MgFe}]^{\prime}$}

\def\kms{$\rm km\;s^{-1}$}

\def\eg{{\it e.g.},}

\def\ie{{\it i.e.},}

\def\apj{ApJ}
\def\aj{AJ}
\def\apjl{ApJL}
\def\apjs{ApJS}
\def\mnras{MNRAS}
\def\aaps{A\&AS}
\def\aap{A\&A}

\def\spose#1{\hbox to 0pt{#1\hss}}

\begin{document}

\title[NGC 4458 and NGC 4478]{Nuclear  stellar discs in
low-luminosity elliptical galaxies: NGC 4458 and NGC 4478}


\author[L. Morelli et al.]{L.~Morelli$^{1,2}$,
  C.~Halliday$^{3}$, E.~M.~Corsini$^{1}$, A.~Pizzella$^{1}$,
  D.~Thomas$^{4}$, R.~P.~Saglia$^{4}$,
\newauthor R.~L.~Davies$^{5}$, R.~Bender$^{4, 6}$, M.~Birkinshaw$^{7}$,
  and F.~Bertola$^{1}$\\
$^1$ Dipartimento di Astronomia, Universit\`a di Padova,
  vicolo dell'Osservatorio~2, I-35122 Padova, Italy.\\
$^2$ European Southern Observatory, 3107 Alonso de Cordova,
  Santiago, Chile.\\
$^3$ INAF-Osservatorio Astronomico di Padova,
  vicolo dell'Osservatorio 5, I-35122 Padova, Italy.\\
$^4$ Max-Planck Institut f\"ur extraterrestrische Physik,
  Giessenbachstrasse, D-85748 Garching, Germany.\\
$^5$ Department of Astrophysics, University of Oxford,
  Keble Road, Oxford, OX1 3RH.\\
$^6$ Universit\"as-Sternwarte, Scheinerstrasse 1, D-81679
  Muenchen, Germany.\\
$^7$ H. H. Wills Physics Laboratory, University of Bristol,
  Tyndall Avenue, Bristol BS8 1TL, UK.\\
}

\date{Received 2004 May 20; accepted 2004 July 21}

\maketitle

\begin{abstract}
We present the detection of nuclear stellar discs in the
low-luminosity elliptical galaxies NGC 4458 and NGC 4478, which are
known to host a kinematically-decoupled core. Using archival HST
imaging, and available absorption line-strength index data based
on ground-based spectroscopy, we investigate the photometric
parameters and the properties of the stellar populations of these
central structures. Their scale length, $h$, and face-on central
surface brightness, $\mu_0^c$, fit on $\mu_0^c$--$h$ relation for
galaxy discs. For NGC 4458 these parameters are typical for nuclear
discs, while the same quantities for NGC 4478 lie between those of
nuclear discs and the discs of discy ellipticals. We present
Lick/IDS absorption line-strength measurements of \Hb,
\Mgb, \Fe\ along the major and minor axes of the galaxies. We model
these data with simple stellar populations that account for the \aFe\/ overabundance. The counter-rotating central disc of NGC 4458 is found
to have similar properties to the decoupled cores of bright
ellipticals. This galaxy has been found to be uniformly old despite
being counter-rotating. In contrast, the cold central disc of NGC 4478
is younger, richer in metals and less overabundant than the main body
of the galaxy. This points to a prolonged star formation history,
typical of an undisturbed disc-like, gas-rich (possibly pre-enriched)
structure.
\end{abstract}


\section{Introduction}
\label{sec:introduction}

In recent years, the sub-arcsec resolution of the {\it Hubble Space
Telescope\/} has allowed the study of galactic nuclei, unveiling the
presence of distinct components such as small-scale stellar discs (see
Pizzella et al. 2002 and references therein) and nuclear clusters
(Carollo et al. 1997; B\"oker et al. 2002; {  Graham \& Guzman 2003}).
To date surface-brightness distributions of nuclear stellar discs
(hereinafter NSD) have been measured for only a few of S0s (Scorza
\& van den Bosch 1998; van den Bosch et al. 1998; Kormendy et
al. 1996) and early-type spiral galaxies (Pizzella et al. 2002). They
have smaller scale lengths ($\sim$10--30 pc) and higher central
surface brightnesses ($\sim$15--19 mag arcsec$^{-2}$ in the $V$ band)
with respect to those of embedded stellar discs, which have been found
in several elliptical and lenticular galaxies (\eg\ Scorza \& Bender
1995, hereinafter SB95).
Although only few NSDs have been studied in detail, they may be quite
a common structure in the central regions of spheroids. {  Indeed,
nuclear discs of gas, dust and stars have been clearly detected in a
large number of early-type galaxies (\eg\ Jaffe et al. 1994; Lauer et
al. 1995; van Dokkum \& Franx 1995; Faber et al. 1997; Carollo et
al. 1997; Tomita et al. 2000; Kormendy et al. 2001; Tran et al. 2001;
Trujillo et al. 2004), but the photometric parameters of such discs
have not yet been derived}.
Furthermore, HST imaging in near-infrared bandpasses provided indirect
signatures of the presence of NSDs as discy isophotes in the nuclei of
early-type galaxies (Ravindranath et al. 2001) or
photometrically-distinct exponential components in bulges (Balcells et
al. 2003, {  2004}).
The existence of NSDs suggests that the continuity of the disc
properties, with a smooth variation of scale parameters from spirals
to discy ellipticals along a sequence of decreasing disc-to-bulge
ratio (Kormendy \& Bender 1996), could be extended to nuclear scales
(van den Bosch 1998). In this framework NSDs could provide important
clues to the assembly scenario of their host galaxies.

In the current picture, the NSDs observed in bulges of S0s and spiral
galaxies are believed to have formed either from a secular evolution
of a nuclear bar (\eg\ NGC 4570, van den Bosch et al. 1998; Scorza \&
van den Bosch 1998; van den Bosch \& Emsellem 1998) or as the end
result of a merging event (\eg\ NGC 4698, Bertola et al. 1999;
Pizzella et al. 2002).
Each of these scenarios is likely to be correct for some but not for
all the objects. In both of them the gas is efficiently directed
toward the galaxy centre, where it first dissipates and settles onto
an equilibrium plane and then forms into stars.  Although the
decoupled kinematics strongly suggest a later infall (\eg\ NGC 4486A,
Kormendy et al. 2001) or merger (\eg\ Holley-Bockelmann \& Richstone
2000), the frequently indistinguishable star formation history remains
an enigma. Yet, these processes are expected to lead to distinct
features in stellar populations of NSDs and surrounding spheroids.
The measurement of the kinematics of some early-type galaxies
(elliptical and S0 galaxies) revealed evidence for kinematically
decoupled or peculiar galaxy cores ({  see Bertola \& Corsini 1999
for a review}), strongly indicative of formation by galaxy-galaxy
merging (\eg\ Kormendy 1984; see Mehlert et al. 1998) and references
therein). Such kinematically-distinct cores have been associated with
a central disc component (\eg\ Rix \& White 1992; Mehlert et
al. 1998).  {  However, triaxial galaxies with radius-dependent
ellipsoids, can, in projection, mimic kinematically decoupled cores
(see Arnold, de Zeeuw \& Hunter 1994, and references therein)}.
Spectroscopic absorption line-strength indices (Faber et al. 1985;
Trager et al. 1998) constrain the luminosity-weighted stellar
population age, metallicity and alpha-element overabundance by
comparison with evolutionary population synthesis models (Worthey
1994; Maraston 1998; Tantalo et al 1998; Vazdekis 1999; Thomas,
Maraston \& Bender 2003, hereafter TMB; Bruzual \& Charlot
2003). Absorption line-strength indices have been measured out to one
effective radii for elliptical galaxies (\eg\ Davies et al. 1993;
Gonz\'{a}lez 1993; Halliday 1999; Mehlert et al. 2000). Such
measurements can distinguish between merging and dissipational
formation scenarios by the constraint of metallicity gradients, dating
of the latest episode of star formation and the measurement of the
timescale over which the bulk of the star formation has taken
place. Davies et al. (1993), Mehlert et al. (2003) and the analysis of
literature data by Kobayashi \& Arimoto (1999) have found elliptical
galaxies to have gradients in metallicity shallower than predicted by
dissipational collapse models.
Mehlert et al. (1998) presented spectroscopic line-strength data and
HST archive imaging for two Coma cluster galaxies with
kinematically-decoupled cores and found little evidence that the cores
have experienced different star formation histories from their parent
galaxies. Using wide-field spectroscopic maps of NGC 4365 acquired
using the SAURON instrument, Davies et al. (2001) presented evidence
for a metal-rich core but constant age and
\aFe-enhancement for both the kinematically-distinct core and
parent galaxy. Both studies are consistent with the core and the
main galaxy having experienced similar star formation
histories.

In this paper, we use archival HST imaging and ground-based
spectroscopy to investigate the photometric parameters and stellar
population diagnostics of the NSDs hosted by the two Virgo galaxies
NGC 4458 and NGC 4478. 
NGC 4458 and NGC 4478 are classified as E0-1 (de Vaucouleurs et
al. 1991, hereinafter RC3) and E2 (Sandage \& Tammann 1981; RC3),
respectively. They are both low-luminosity ellipticals with a
$M_B^0\sim-18$ (RC3) at a distance of 12.6 Mpc (Tully 1988, $H_0=100$
\kms\ Mpc$^{-1}$). 
They have a power-law central luminosity profile ({  Ferrarese et
al. 1994}; Faber et al. 1997; Rest et al. 2001) and both galaxies show
deviations from the $r^{1/4}$ law in their outskirts (Michard 1985;
Prugniel et al. 1987; Caon et al. 1990; Peletier et al. 1990), which
have been explained as due to tidal interaction with NGC 4461 and NGC
4486.
Recently, the stellar kinematics along the major and minor axes of NGC
4458 and NGC 4478 have been measured by Halliday et al. (2001), who
found the signature of a kinematically-decoupled core in their inner
$\sim5$ arcsec. In particular, NGC 4458 has a clear counter-rotating
core along the major axis, while NGC 4478 has a cold component
detected along both the major and the minor axis.

In Section \ref{sec:photometry} we perform the photometric analysis of
the archival HST images of the nuclei of NGC 4458 and NGC 4478
in order to derive the surface brightness profiles and scale
parameters of their NSDs. In Section \ref{sec:spectroscopy} we
present the absorption line-strength indices as a function of radius
of the major and minor axes of both galaxies from the thesis research
of Halliday (1999). In Section
\ref{sec:agemetover} we combine photometric and spectroscopic
results to assess age, metallicity and overabundances of stellar
populations of the two NSDs and surrounding spheroids. This allowed
us to suggest possible formation and evolution scenarios for NGC 4458
and NGC 4478. We summarize our conclusions in Section \ref{sec:conclusions}.

\section{Photometric parameters}
\label{sec:photometry}

\subsection{Data reduction}
\label{sec:data_reduction}

We retrieved Wide Field Planetary Camera 2 (WFPC2) images of NGC 4458
and NGC 4478 from the HST archive. Data for the filter F814W were
selected as a compromise between obtaining data in identical
bandpasses for both galaxies and minimizing the effects of dust on
photometric measurements.
Total exposure times were 1120 s for NGC 4458 (5 exposures,
Prog. Id. 5512, P.I. S.~M.~Faber) and 1600 s for NGC 4478 (2
exposures, Prog. Id. 6587, P.I. D. Richstone), respectively.
All exposures were taken with the telescope guiding in fine lock,
which typically gave an rms tracking error of $0.003$ arcsec.  We
focused our attention on the Planetary Camera chip (PC) where the
nucleus of both galaxies was centred. This consists of $800\times800$
pixels of $0.0455\times0.0455$ arcsec$^2$ each, yielding a field of
view of about $36\times36$ arcsec$^2$.
The images were calibrated using the standard reduction pipeline
maintained by the Space Telescope Science Institute. Reduction steps
include bias subtraction, dark current subtraction, and flat-fielding
and are described in detail in Holtzman et al. (1995a).
Subsequent reduction was completed using standard tasks in the {\tt
STSDAS} package of {\tt IRAF}\footnote{{\tt IRAF} is distributed by
NOAO, which is operated by AURA Inc., under contract with the National
Science Foundation}. Bad pixels were corrected by means of a linear
one-dimensional interpolation using the data quality files and the
{\tt WFIXUP} task.
Different images of the same target were aligned and combined using
{\tt IMSHIFT} and knowledge of the offset shifts. Cosmic ray events
were removed using the task {\tt CRREJ}.
The cosmic-ray removal and bad pixel correction were checked by
inspection of the residual images between the cleaned and combined
image and each of the original frames. Residual cosmic rays and bad
pixels in the PC were corrected by manually editing the combined image
with {\tt IMEDIT}.
The sky level ($\sim1$ count pixel$^{-1}$) was determined from regions free
of sources in the Wide Field chips and subtracted from the PC frame
after appropriate scaling.

\subsection{Detection of nuclear discs}
\label{sec:detection}

\begin{figure}
   \epsfig{figure=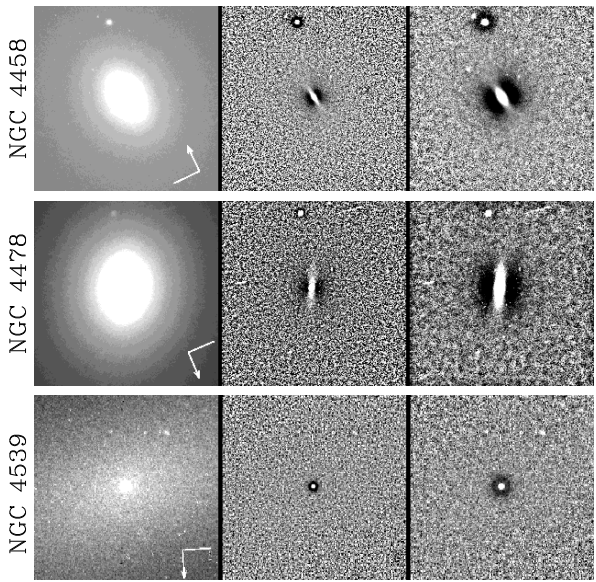,width=8.5truecm}
\caption[]{{\it Left panels:\/} WFPC2/F814W images of NGC 4458
  and NGC 4478 (for which nuclear  discs were detected), and NGC 4539
  (for which a nuclear  disc was not detected). The size of the plotted
  region is $19.3\times19.3$ arcsec$^2$. The orientation is specified
  by the arrow indicating north and the segment indicating east in the
  lower right corner of each panel. {\it Middle and right panels:\/}
  Unsharp masking of the WFPC2/F814W images obtained with $\sigma=2$
  and $6$ pixels, respectively. Sizes and orientations are as in the
  left-hand panels.}
\label{fig:unsharp}
\end{figure}

To test for the presence of a NSD in NGC 4458 and NGC
4478, we constructed the unsharp-masked image of the PC frame using an
identical procedure to Pizzella et al. (2002). We divided each image
by itself after convolution by a circular Gaussian of width $\sigma=2$
and 6 pixels, corresponding to $0.09$ arcsec and $0.27$ arcsec,
respectively (Figure \ref{fig:unsharp}).
This procedure enhanced any surface-brightness fluctuation and
non-circular structure extending over a spatial region comparable to
the $\sigma$ of the smoothing Gaussian. Two values of $\sigma$ were
adopted to attempt the identification of structures with different
scale lengths.

\begin{figure}
  \epsfig{figure=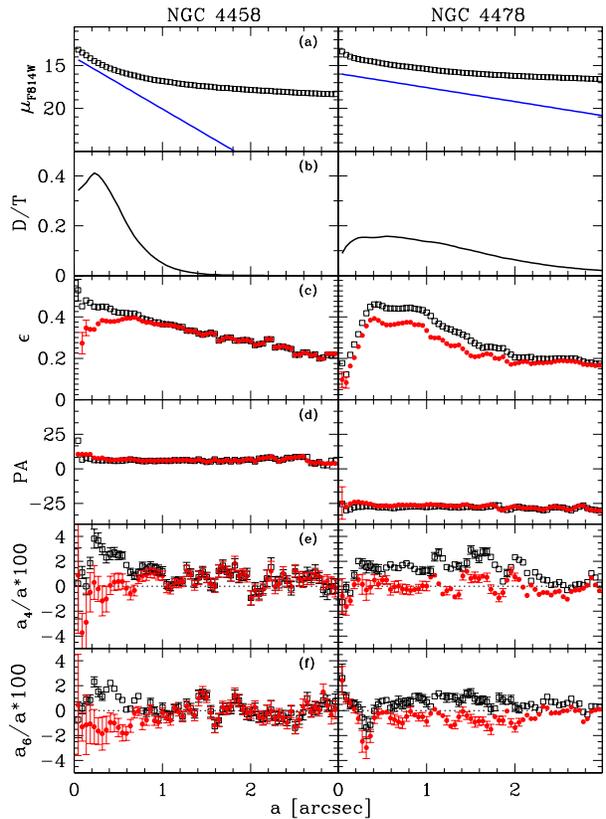,width=8.5truecm}
\caption{{\it (a)\/} Radial profiles of surface brightness of NGC 4458 ({\it
  left panel\/}) and NGC 4478 ({\it right panel\/}) after
  deconvolution ({\it open squares\/}) are compared to those of their
  nuclear discs ({\it thick lines\/}). {  {\it (b)} Radial profiles of
  the fraction of total luminosity contributed by the nuclear disc of
  NGC 4458 ({\it left panel\/}) and NGC 4478 ({\it right panel\/}).}
  Radial profiles of {\it (c)\/} ellipticity, {\it (d)\/} position
  angle, {\it (e)\/} $a_4$ and {\it (f)\/} $a_6$ Fourier coefficients
  before ({\it open squares\/}) and after ({\it filled circles\/}) the
  subtraction of the best-fitting model for the nuclear disc of NGC
  4458 ({\it left panels\/}) and NGC 4478 ({\it right panels\/}).}
\label{fig:photometry_disk_subtracted}
\end{figure}

A highly elongated structure is clearly visible in the nucleus of each
galaxies.
These structures are not artifacts of the unsharp-masking
procedure. In Figure \ref{fig:unsharp} we show as a counter-example the
unsharp-masked image of NGC 4539 which has been obtained using an
identical method to that for NGC 4458 and NGC 4478; we clearly do not
detect an elongated disc-like structure. Moreover the nuclear
structures of NGC 4458 and NGC 4478 are associated with a central
increase in ellipticity measured by performing an isophotal analysis
using the {\tt IRAF} task {\tt ELLIPSE}
(Figure \ref{fig:photometry_disk_subtracted}).

We analyzed the isophotal profiles of both galaxies by first masking
foreground stars and then fitting ellipses to the isophotes. We
allowed the centres of the ellipses to vary, to test whether the
galaxies were disturbed. Within the errors of the fits, we found no
evidence of variations in the fitted centre. The ellipse fitting was
repeated with the ellipse centres fixed. The resulting azimuthally
averaged surface brightness, ellipticity, position angle, the fourth
($a_4$), and sixth ($a_6$) cosine Fourier coefficients profiles are
presented in Figure \ref{fig:photometry_disk_subtracted}.
In the innermost $\sim2$ arcsec we measured positive values of the
$a_4$, and $a_6$ Fourier coefficients, which describe the discy
deviation of the isophotes from pure ellipses {  (Jedrzejewski
1987)}. These photometric features have been also observed in the
HST/F555W image for NGC 4458 (Lauer et al. 1995; {  Trujillo et
al. 2004}) and in the HST/F702W image of NGC 4478 (van den Bosch et
al. 1994; Rest et al. 2001; {  Trujillo et al. 2004}) and confirm
the presence of a NSD in both galaxies.

\subsection{Photometric decomposition}
\label{sec:decomposition}

\subsubsection{Photometric parameters of the nuclear discs
with the Scorza \& Bender method}

\begin{figure*}
   \epsfig{figure=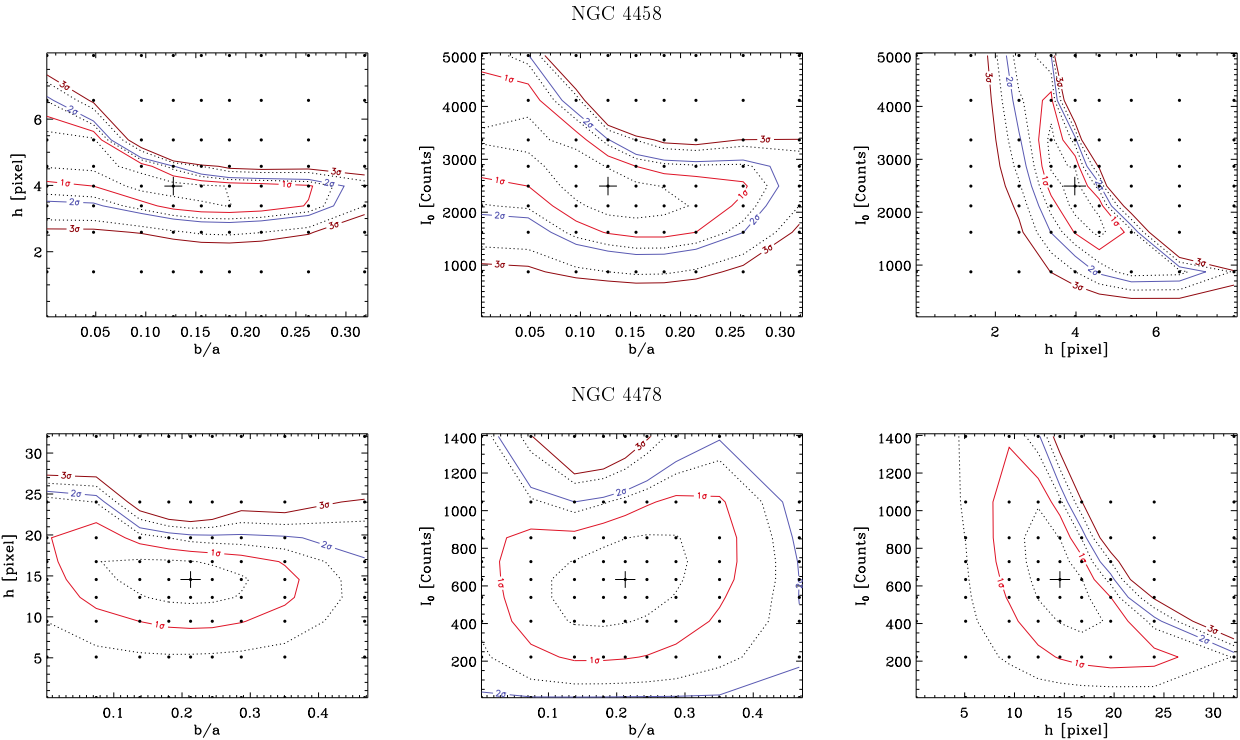,width=10truecm,angle=270}
\caption{Contour plots of $\chi_\nu^2$ for
  NGC 4458 ({\it upper panels\/}) and NGC 4478 ({\it lower panels\/})
  as a function of the central surface brightness, $I_0$, scale
  length, $h$, and axial ratio, $b/a$, of the nuclear disc models.
  {\it Solid contours\/} define the formal $68.3\%$, $95.4\%$, and
  $99.7\%$ confidence levels. Intermediate confidence levels are given
  with {\it dashed contours\/}.  {\it Filled circles\/} indicate
  actual model calculations. Bi-cubic interpolation is used to
  calculate at $\chi_\nu^2$ intermediate points.  {\it Crosses\/}
  correspond to the best-fitting photometric parameters of the nuclear
  disc.}
\label{fig:confidence_levels}
\end{figure*}

After establishing the existence of NSDs, we measured their
photometric properties using the method described by SB95. This
method is based on the assumption that isophotal disciness is the
result of the superimposition of a spheroidal component (which is
either a host elliptical galaxy or a bulge component) and an inclined
disc. The two components are assumed to have both perfectly elliptical
isophotes with constant but different ellipticities.
When adopting this technique to study the innermost regions of
galaxies it is essential to restore the images from the effects of the
HST point spread function (PSF) in order to properly derive
the photometric parameters of the NSDs, as shown by Scorza \&
van den Bosch (1998). Such deconvolution was performed through the
Richardson-Lucy method by means of the {\tt IRAF} task {\tt
LUCY}. Although susceptible to noise amplification, this algorithm was
shown by van den Bosch et al. (1998) to lead to a restored
surface-brightness distribution comparable to that obtained using a
multi-Gaussian representation (Monnet et al.  1992; Cappellari
2002).
We believe that the results obtained by van den Bosch et al. (1998)
are directly applicable to our case: we are dealing with images
obtained with similar or longer integration times, galaxies with less
steep surface-brightness profiles and NSDs with equal or
larger scale lengths.
We decided to deconvolve the images with a number of iterations
between 3 and 6. A larger number of iterations does not affect the
result of the decomposition but does amplify the noise.
For each given image and nucleus position on the PC we adopted a model
PSF calculated using the {\tt TINY-TIM} package (Krist \& Hook 1999).
No correction for telescope jitter was necessary.
The SB95 method consists of the iterative subtraction of a thin disc
model characterized by an exponential surface-brightness profile with
central surface brightness $I_0$ and radial scale length $h$, and by
an axial ratio $b/a$. The disc parameters are adjusted until the
departures from perfect ellipses are minimized (\ie\ $a_{4}$ and
$a_{6}$ are close to zero).
For each disc model we obtained the disc-free image of the galaxy by
subtracting the disc model from the galaxy image. We performed an
isophotal analysis on the disc-free image using the {\tt IRAF} task
{\tt ELLIPSE}. We defined

\begin{equation}
\chi^2 =\sum_{i=1}^{N} \frac{a^2_{\it 4,disc-free}(i)}{\sigma(i)^2}
\label{eq:chi}
\end{equation}
where $a_{\it 4,disc-free}(i) \pm \sigma(i)$ is the value of the $a_4$
Fourier coefficient measured for the $i-$th isophote in the disc-free
image, and $N$ is the number of fitted isophotes. We assumed
$\sigma(i)=0.01$ as typical error on $a_{\it 4,disc-free}$ for all the
isophotes in the region of the NSD. We defined the reduced $\chi^2$ as
$\chi_\nu^2=\chi^2/(N-M)$, where $M=3$ is the number of free
parameters, namely $I_0$, $h$, and $b/a$. According to our technique
for photometric decomposition the minimum value of
$\chi_\nu^2\equiv\chi_{\nu,{\it min}}^2$ corresponds to the
best-fitting model of the NSD. {  We found $\chi_{\nu,{\it
min}}^2=0.43$ for NGC 4458 and $\chi_{\nu,{\it min}}^2=0.34$ for NGC
4478.} We determined $\Delta\chi_\nu^2\equiv\chi_\nu^2-\chi_{\nu,{\it
min}}^2$ and derived its confidence levels under the assumption that
the errors are normally distributed (Press et al. 1992). The
resulting contour plots of $\chi_\nu^2$ are shown in Figure
\ref{fig:confidence_levels} and they allowed us to derive the
best-fitting values of $I_0$, $h$ and $b/a$ and their $1\sigma$
errors.

We derived the Johnson $V-$band central surface brightness of the
NSDs from the HST VEGAMAG system (Holtzman et
al. 1995b) using the {\tt STSDAS} task {\tt SYNPHOT}. Since this
correction depends on the galaxy spectral energy distribution, it has
been calculated using the spectral template for elliptical galaxies by
Kinney et al. (1996). The resulting relation is $V-m_{\rm F814W}=1.33$.
We derived the scale length of the NSDs assuming a distance
of 12.6 Mpc for both galaxies.
The inclination of NSDs has been calculated as
$i=\arccos{(b/a)}$.
The resulting values of central surface brightness in the $V$ band,
scale length and inclination for both NSDs are listed in
Table \ref{tab:disk_parameters}.

The disciness of NGC 4458 and NGC 4478 disappears after the
subtraction of the best-fitting model of the NSD as it results
from the elliptical shape of their isophotes as shown by the
photometric profiles plotted in Figure
\ref{fig:photometry_disk_subtracted}.
We tested that the result of the photometric decomposition is
independent of position on the PC chip where the PSF is built.  { 
The resulting fraction of total luminosity contributed by the nuclear
disc of NGC 4458 and NGC 4478 is plotted as a function of radius in
Figure \ref{fig:photometry_disk_subtracted}.}

\subsubsection{Photometric parameters of the nuclear discs
with the Seifert \& Scorza method}

In applying the decomposition method by SB95 we assumed that the
surface-brightness profile of the NSDs was exponential. To
verify this assumption, we derived the surface-brightness profile of
the NSDs using the alternative method of Seifert \& Scorza
(1996, hereinafter SS96). This decomposition method was developed to
identify embedded discs in ellipticals and bulges without any {\it a
priori\/} assumptions of the parametric law for their
surface-brightness profile.
In the deconvolved image of the galaxy we masked the region of the
NSD with a double-cone mask, and fit ellipses to the
isophotes in the remaining part of image using the {\tt IRAF} task
{\tt ELLIPSE} to derive the photometric properties of the host galaxy.
A double-cone mask centred on the galaxy major axis is an optimal
choice for the highly-inclined NSDs of NGC 4458 and NGC 4478
since we expect that the isophotes close to the galaxy minor axis will
be only slightly contaminated by light from the NSDs. The
very central portion of the galaxy image was not masked to allow
meaningful isophotes to be fitted. These masks correspond to the
central $3\times3$ pixel ($0.14\times0.14$ arcsec$^2$) of NGC 4458 and
$5\times5$ pixel ($0.23\times0.23$ arcsec$^2$) of NGC 4478.
The opening angle of the mask ($\approx90^\circ$) was adjusted to
minimize the influence of the NSD on the fitted parameters of
the surrounding galaxy bulge component, but to ensure that a
sufficient number of data points for the model derivation remained. A
model was constructed to have perfectly elliptical isophotes with the
same surface brightness, ellipticity and position angle radial
profiles which we obtained from the isophotal fit. 
We subtracted the galaxy model from the original image to obtain a
residual image which shows an elongated structure, namely
the NSD, responsible for the higher-order deviations from
perfectly elliptical isophotes.
We extracted the surface-brightness radial profile along the major
axis of the residual disc which was close to being exponential. Scale
lengths are consistent within their errors to those measured using the
SB95 method, while the central surface brightnesses are somewhat
fainter.
This discrepancy in the surface brightnesses derived for using the
SB95 and SS96 methods is due to the actual shape of the
double-cone mask where we excluded the central region in order to
allow meaningful isophotes to be fitted. In this way the observed
central surface brightness is assigned to the host galaxy and the
central surface brightness of the NSD is therefore
underestimated.

\begin{table}
\caption{Photometric parameters of the nuclear discs}
\label{tab:disk_parameters}
\begin{tabular}{ccccc}
\hline
\noalign{\smallskip}
\multicolumn{1}{c}{Name} &
\multicolumn{1}{c}{$\mu_{0,V}$} &
\multicolumn{1}{c}{$h$} &
\multicolumn{1}{c}{$i$} &
\multicolumn{1}{c}{$L_{\it disc}$} \\
\multicolumn{1}{c}{} &
\multicolumn{1}{c}{[mag arcsec$^{-2}$]} &
\multicolumn{1}{c}{[pc]} &
\multicolumn{1}{c}{[$^\circ$]} &
\multicolumn{1}{c}{[L$_{\it \odot,V}$]} \\
\noalign{\smallskip}
\hline
\noalign{\smallskip}
NGC 4458 & $14.74_{-0.67}^{+0.71}$ & $11.0_{-2.5}^{+5.9}$ & $83.1_{-8.8}^{+6.9}$ & $4.8\cdot10^6$\\
\noalign{\smallskip}
NGC 4478 & $16.61_{-0.81}^{+1.48}$ & $40.5_{-18.7}^{+33.0}$ & $77.9_{-10.0}^{+11.6}$ & $1.9\cdot10^7$\\
\noalign{\smallskip}
\hline
\end{tabular}
\end{table}

\subsection{Structural parameters of nuclear discs}
\label{sec:parnucdic}

It is interesting to compare the scale lengths given in Table
\ref{tab:disk_parameters} ($h\approx 0.2''$ for NGC 4458 and $h\approx
0.7''$ for NGC 4478)
with the ``break'' radii found in the major and minor axis kinematics
of the two galaxies (Halliday et al. 2001).  Counter-rotation of about
(30 \kms) is observed in the inner 4 arcsec of the major axis of NGC
4458.  In the case of NGC 4478 the velocity dispersion drops in the
central 2 arcsec, where the rotational velocity along the major axis
drops by $\approx 20$ \kms\ from 50 \kms\ at 2 arcsec to 30 \kms\ at 3
arcsec from the centre to increase again at larger radii.  

Clearly, the spatial resolution of the kinematic data {  (see Section
\ref{sec:spectroscopy} for details)}
is smearing the signal of the NSDs.
Figure 4 shows the position of the NSDs of NGC 4458 and NGC 4478 in
the disc $\mu_0^c$--$h$ diagram (adapted from Pizzella et al. 2002;
{  but see also van den Bosch 1998, and Graham 2001}).  The face-on
central surface brightness $\mu_0^c$ have been derived from the
observed one $\mu_0$ by applying the following inclination correction
\begin{equation}
\mu_0^c\,=\,\mu_0\,-2.5\,\log({\cos{i}}) 
\end{equation}
and without taking into account any correction for extinction.
The (small) scale length and (high) central surface brightness of the
disc of NGC 4458 are typical for NSDs.  In contrast, the (relatively
large) scale length and (relatively low) central surface brightness of
the disc of NGC 4478 are in between the typical values found for discy
ellipticals and NSDs. No correlations are found with global properties
(\eg\ luminosity, scale-length, disc-to-bulge ratio) of the host
galaxies.

{  Finally, on a speculative line of thought without presumption of
rigourness, it is interesting to correlate the order of magnitude
rotational velocities estimated above (30 \kms\ for NGC 4458 and 50
\kms\ for NGC 4478)} with the estimated magnitudes of the discs. We
find that discs following the Tully-Fisher relation (as in Haynes et
al. 1999) and having the same rotational velocities, are expected to
be $\approx 3$ mag brighter than the NSDs of NGC 4458 and NGC 4478. In
other words, the NSDs of NGC 4458 and NGC 4478 rotate faster than
discs of the same luminosity that follow the Tully-Fisher relation.
Taken at face value, this might indicate that they behave like test
particles embedded in the dominant gravitational potential generated
by the main (bulge) body of the galaxies. However, the situation is
complicated by the low spatial resolution of the kinematic data. Using
the total masses of the discs derived in Section \ref{sec:agemetover},
we estimate that maximum rotational velocities of the order of
$\approx 30-50$ \kms\ are generated by the NSDs alone around 0.6-2
arcsec. Clearly, higher resolution kinematic data {  and proper
modelling (including a decomposition of the line-of-sight velocity
distributions and seeing convolution) are needed to pin down the
issue.}

\begin{figure}
\begin{center}
\begin{minipage}{3in}
    \epsfig{figure=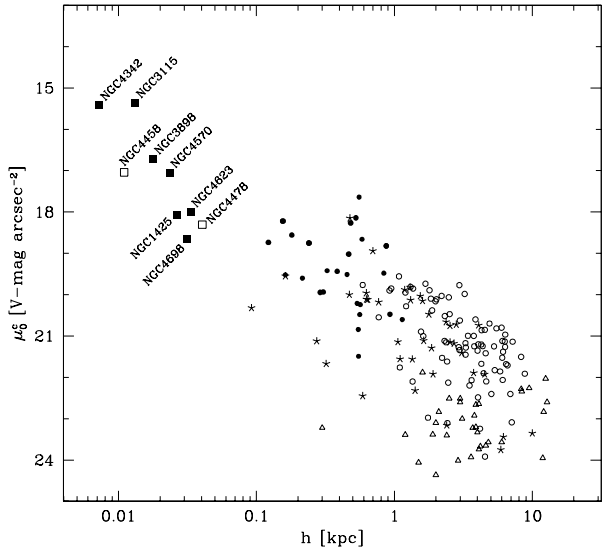,width=8truecm}
\caption{Disc $\mu_0^c$--$h$ diagram adapted from Pizzella et al. (2002).
  {\it Open circles\/} refer to high surface-brightness spirals, {\it
  triangles\/} to low surface-brightness spirals, {\it stars\/} to
  S0's, {\it filled circles\/} to discy ellipticals. {\it Large
  squares\/} correspond to NSDs in elliptical galaxies ({\it open
  symbols}) and in S0's and spiral galaxies ({\it filled symbols}),
  respectively.}
\end{minipage}
\label{fig:npar}
\end{center}
\end{figure}

\section{Line-strength indices}
\label{sec:spectroscopy}

Absorption line-strength indices are taken from a study of the stellar
populations and kinematics of low-luminosity elliptical galaxies, in
the Virgo cluster and nearby groups (Halliday 1999; Halliday et al.
2001). {  We provide a brief summary of our spectroscopic
observations. Long-slit spectroscopic data were obtained using the
Blue Channel Spectrograph at the Multiple Mirror Telescope, Arizona
(USA). A Loral 3k$\times$1k CCD chip with a pixel size of
15$\mu$m$\times$15$\mu$m, a grating of 1200 slits mm$^{-1}$ and a slit
width of 1~arcsec were used. The spectral resolution was
$\sim$1.5~\AA\ FWHM, wavelength range 4555 - 6045~\AA\ and spatial
scale 0.6 arcsec pixel$^{-1}$. Seeing varied between 0.5 and
0.8~arcsec during observations. The reader is referred to Halliday
(1999) for details of the spectroscopic data reduction and Halliday et
al. (2001) for a description of the kinematical analyses.}

\subsection{Measurement of absorption line-strength indices}
\label{sec:indices_measurement}

Line-strength indices were measured as a function of radius to
approximately one photometric effective radius along the major and
minor axes of each galaxy. One-dimensional spectra were extracted for
different luminosity-weighted radii of each axis and line-strength
indices were measured for each spectrum using an algorithm developed
by G. Baggley and revised by H. Kuntschner (Halliday 1999). The
Lick/IDS system line-strength index definitions of Trager et
al. (1998) were adopted. A fuller description of the algorithm is
given below.

The principal input was a spectroscopic CCD frame for which a basic
data reduction had been completed. Before extracting one-dimensional
spectra the galaxy spectrum was checked to be aligned with the rows of
the CCD frame. Using the {\tt IRAF} tasks {\tt APFIND} and {\tt
APTRACE} the position of the galaxy centre was traced along the
dispersion axis; this step was performed to an accuracy of
$\simeq$0.03 pixels rms. The pixels were resampled such that the
galaxy spectrum was centred on the same row pixel position as a
function of dispersion.

One-dimensional spectra were extracted by summing rows of the galaxy
frame until a signal-to-noise ratio (hereinafter S/N) of 40 per \AA\
(for the wavelength range $\sim$5100--5300 \AA) was attained. The
dispersion axes of extracted spectra were rebinned to logarithmic
intervals of wavelength. Redshifts were measured using the {\tt IRAF}
task {\tt FXCOR}, by cross-correlating each spectrum with a range of
one-dimensional de-redshifted stellar spectra. The mean redshift
determined for all templates was used to de-redshift each galaxy
spectrum. H$\beta$, \Mgb , Fe5270 and Fe5335 line-strength indices
were measured for all spectra. Errors for these raw index measurements
were calculated based on Poissonian noise, the typical sky level
counts subtracted during sky subtraction, and the values of read-out
noise and gain of the CCD.

Our line-strength index measurements were transformed to the Lick/IDS
system. This procedure involved two steps: (i) correction to zero
galaxy velocity dispersion and (ii) correction to the Lick/IDS
spectral resolution ($\sim$8.6 \AA).

To correct for the effects of galaxy velocity dispersion, $\sigma$,
different stellar spectra were smoothed by Gaussians corresponding to
measurements of $\sigma$ between 0 km s$^{-1}$ and 300 km s$^{-1}$ in
intervals of 20 km s$^{-1}$. Smoothing was performed using the {\tt
IRAF} task {\tt GAUSS}. Correction factors for different values of
$\sigma$ were calculated for each index by comparing line index
measurements for each smoothed spectrum with measurements for the
original unsmoothed stellar spectrum. For the atomic indices H$\beta$,
\Mgb, Fe5270 and Fe5335 correction factors were defined to be the
ratio of measurements for the original unbroadened stellar spectrum,
$I_{orig}$, and that for the spectrum broadened to a particular
velocity dispersion, $I_{\sigma}$,
\ie\
\begin{equation}
{C(\sigma) = \frac{I_{orig}}{I_{\sigma}}.}
\end{equation}
\Mgb, Fe5270 and Fe5335 line-strength index measurements
were corrected for the effects of $\sigma$ using the calibrations for
12 stellar observations; {  for H$\beta$ data 8 stellar
observations} were used. Linear relations were interpolated between
the mean correction factor for each $\sigma$. The correction factor
for an arbitrary measurement of $\sigma$ was determined using the
linear relation for the $\sigma$ range bracketing the $\sigma$
measurement. The measurement of $\sigma$ as a function of radius was
presented for both galaxies in Halliday et al. (2001). We adopt the
$\sigma$ measurements obtained for a $\rm S/N\simeq60$ per \AA .
Measurements of $\sigma$ with $S/N = 30$~per \AA\ were adopted at
large radii where uncertainties were larger.

Absorption line-strength index measurements and measurements of
$\sigma$ from Halliday et al. (2001) were measured in general for
different galactic radii. Linear interpolations were made between
neighboring measurements of $\sigma$. A $\sigma$ measurement for a
radius at which line-strength indices were measured, was found by
linear interpolation from bracketing data.  Depending on the
appropriate side of the galaxy, if the radius was either less than or
exceeded the radius of the most reliable measurement of $\sigma$, the
linear interpolation determined for the closest radial interval was
used to determine $\sigma$.  The errors of line-strength index
measurements corrected for galaxy velocity dispersion were calculated
to be,
\begin{equation}
{\delta I_{atomic}(\sigma) = \sqrt{( \delta I_{orig} \cdot C(\sigma) )^2
  + ( I_{orig} \cdot \delta C(\sigma) )^2}}
\end{equation}
where $\delta I_{orig}$ is the error in the raw measurement calculated
as outlined above and $\delta C(\sigma$) is the addition in quadrature
of the standard deviations of correction factors for measurements of
$\sigma$ bracketing the $\sigma$ measurement.

Corrections for the difference between our spectral resolution and the
resolution of the Lick/IDS system were applied. A simulated Lick
aperture was extracted for the major axis spectrum of each galaxy
studied by Halliday (1999) present in the ``pristine'' IDS sample of
Trager (1997). {  Spectra were resampled such that the galaxy centre
was aligned with the same pixel row position as a function of
wavelength}. For our spectroscopic slit width of $1$ arcsec {  and
spatial scale of $0.6$ arcsec pixel$^{-1}$ a} one-dimensional spectrum
corresponding to a Lick/IDS aperture was created by summing the
central 9 rows (\ie\ the central $\sim5.4$ arcsec of each galaxy
spectrum). Line-strength indices were measured using the Lick/IDS
definitions of Trager et al. (1998). The galaxy velocity dispersion
was measured using the {\tt IRAF} task {\tt FXCOR}. Line-strength
indices were corrected for the effects of $\sigma$ using the
calibrations discussed above. Direct comparisons were made with the
fully-corrected Lick/IDS $1.4\times4.0$ arcsec$^2$ aperture
measurements of the ``pristine'' IDS sample of Trager (1997); we { 
considered} the Lick/IDS H$\beta$ measurements corrected for the
effects of $\sigma$ but uncorrected for H$\beta$ emission. Relation
fitting was attempted between the simulated Lick/IDS aperture
measurements corrected for $\sigma$ and the Lick/IDS measurements of
Trager (1997), as a function of our $\sigma$-corrected, simulated
Lick/IDS aperture measurements. After testing various routines, the
median value of offset was found to provide a good fit for all
line-strength indices (\ie\ for H$\beta$, 0.272; Mg$_{2}$, -0.048;
Mg$_{b}$, 0.051; Fe5270, 0.544; Fe5335, 0.697). The $\sigma$-corrected
measurements were established on to the Lick/IDS system by subtracting
these median values for all index measurements excluding \Mgb~{  for
which no zero correction was required due to the large
scatter of offset values around the median.}

\subsection{Absorption line-strength gradients}
\label{sec:indices_gradients}

\begin{figure*}
\begin{center}
 \begin{minipage}[b]{.46\linewidth}
  \centering\epsfig{figure=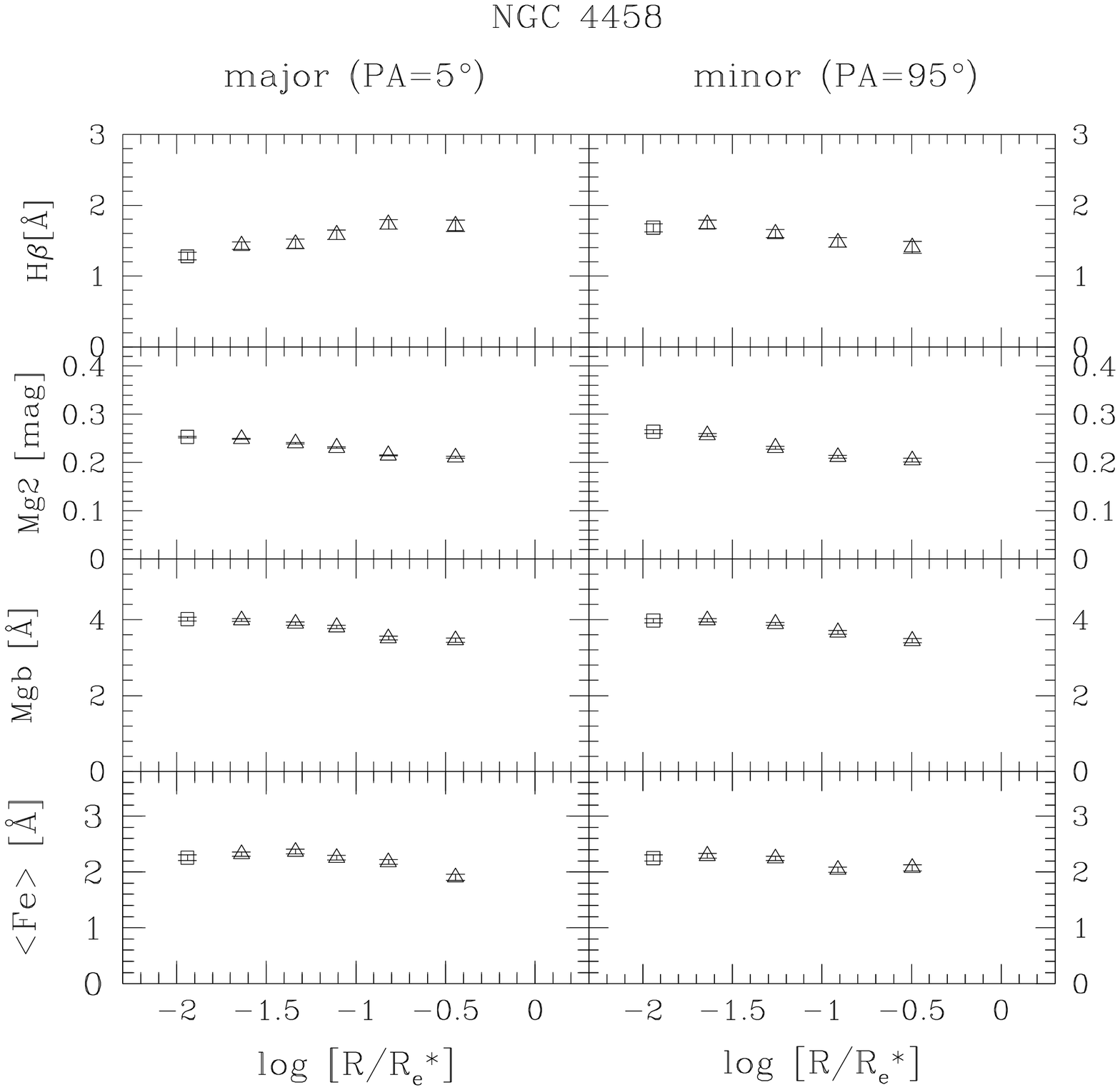,width=\linewidth}
 \end{minipage} \hfill
 \begin{minipage}[b]{.46\linewidth}
  \centering\epsfig{figure=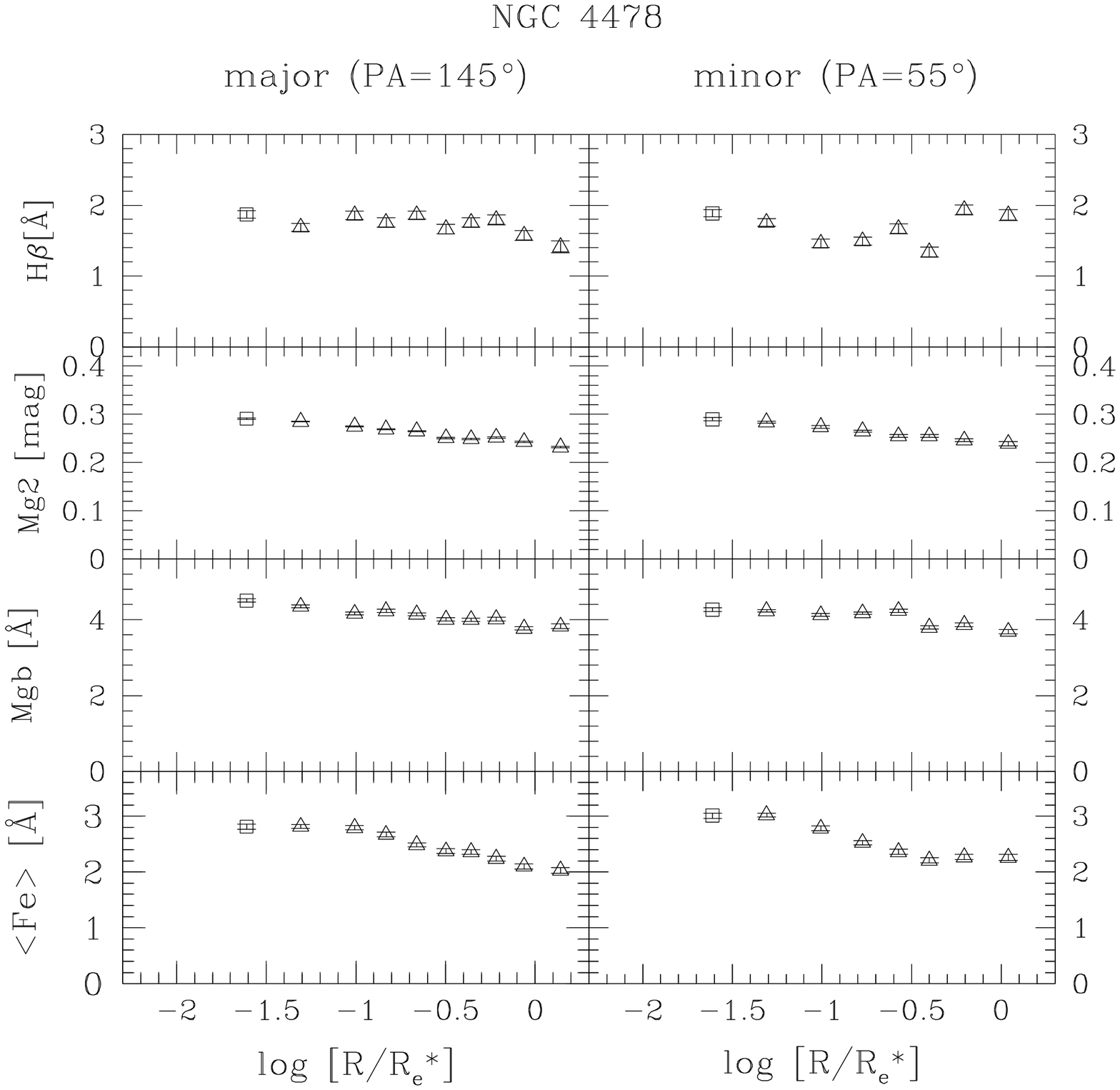,width=\linewidth}
 \end{minipage}
\caption{The radial profiles of the line-strength indices measured
  along the major and minor axis of NGC 4458 ({\it left panels\/}) and
  NGC 4478 ({\it right panels\/}). Squares refer to the value of
  line-strength indices measured at the galaxy centre. 
\label{fig:indices}}
\end{center}
\end{figure*}

Absorption line-strength index measurements for H$\beta$, \Mgb,
Mg$_{2}$ and $\rm{\left<Fe\right> = (Fe5270 + Fe5335)/2}$ are shown as
a function of radius for the major and minor axes of NGC 4458 and NGC
4478 in Figure \ref{fig:indices}. Line-strength measurements are
presented as a function of the normalized radius $R/R_e^*$, where
$R_e^* = R_e/\sqrt{1-\epsilon}$ for the major axis and $R_e^* =
R_e\cdot\sqrt{1-\epsilon}$ for the minor axis, respectively. $R_e$ is
the circularized effective radius defined as $R_e=A_e/2$ with { 
$A_e$ the diameter of the circle enclosing one-half of the $B-$band
total flux (RC3)}, and $\epsilon$ is the mean ellipticity of the
galaxy.
We assume $R_e=26.1$ arcsec (RC3, corresponding to 1.6 kpc at the
adopted distance) and $\epsilon=0.00$ (Lauer et al. 1995) for NGC
4458, $R_e=13.4$ arcsec (RC3, corresponding to 0.8 kpc at the adopted
distance) and $\epsilon=0.17$ (Peletier et al. 1990) for NGC 4478.
The different central values measured along major and minor axis for
the same index give an estimate of the true measurement errors.
Measurements for Mg$_{2}$ are shown for completeness and the
calibration of this molecular index to the Lick/IDS system is described in
Halliday (1999).

\begin{figure}
\begin{center}
\begin{minipage}{4in}
    \epsfig{file=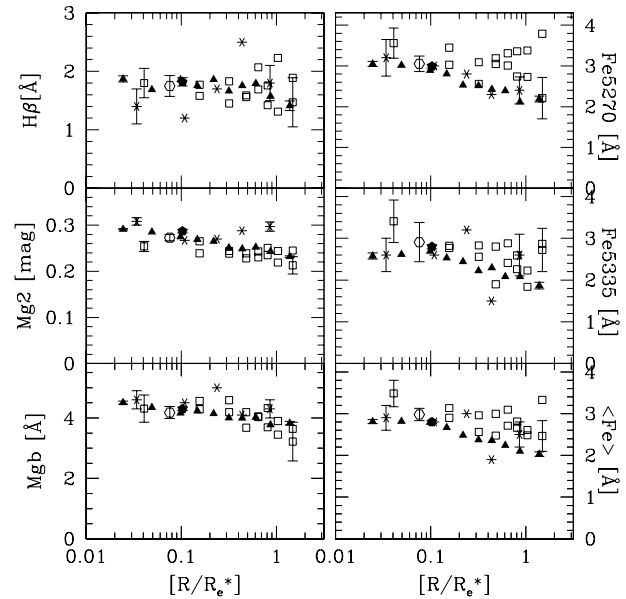,width=8.5cm}
\end{minipage}
\caption{ Comparison of our line-strength index data for NGC 4478
  measured as a function of radius of the major axis (shown by filled
  triangles\/ symbols), the data Gorgas et al. (1990, {\it open
  squares\/}) and Peletier (1989, {\it asterisks\/}). The central
  measurements of Gonz\'{a}lez (1993, {\it pentagon\/}) and Trager et
  al. (1998, {\it hexagon\/}) are shown for representative
  radii. Error bars are provided for data at the innermost and
  outermost radii and the central measurements.
\label{fig:n4478_comparison}}
\end{center}
\end{figure}

For NGC 4478, measurements of H$\beta$, \Mgb, Mg$_2$, Fe5270, Fe5335
and \Fe\ were derived previously for the galaxy centre by Gonz\'{a}lez
(1993) and Trager et al. (1998), and as a function of radius of the
major axis by Peletier (1989) and Gorgas et al. (1990). In Figure
\ref{fig:n4478_comparison} we compare all previous measurements with
our measurements presented here.

Good agreement within all margins of uncertainty is found between our
data and the measurements of Peletier (1989). Measurements and errors
for Fe5270 were not tabulated by Peletier (1989) and have been derived
using the corresponding measurements of Fe5335 and \Fe .
Agreement within the errors is found between our measurements and
those of Gorgas et al. (1990) for H$\beta$ and \Mgb\ but not for
Fe5270, Fe5335, and \Fe. A single bad pixel column was interpolated
for a region corresponding to the red passband of Fe5270 of our
spectrum of NGC 4478 but this is unlikely to be the cause of the
discrepancy here since no evidence of unsuccessful bad pixel
correction was found. Since Fe5270, Fe5335, and \Fe\/ are critical
indices in the study of stellar populations, it has to be noted that
our results are based on our data. At the galaxy centre there is a
small difference between our value of the Mg$_2$ index and that of
Gorgas et al. (1990), while agreement within the errors is found for
data at all other radii.
The central measurements of the line-strength indices of Gonz\'{a}lez
(1993) and Trager et al. (1998) are in good agreement with our
measurements.

\section{Ages, metallicities and \aFe ~overabundances}

\label{sec:agemetover}

In the following we indicate the combined
Magnesium-Iron index with
\begin{equation}
[{\rm MgFe}]^{\prime}=\sqrt{{\rm Mg}\,b(0.72\times\mathrm{Fe5270} +
0.28\times\mathrm{Fe5335})}
\label{eqmgfep}
\end{equation}
newly defined by TMB. This index is completely independent of the
[\aFe]\ overabundance and hence serves best as a metallicity tracer.

\begin{figure*}
\begin{minipage}{\linewidth}
    \epsfig{file=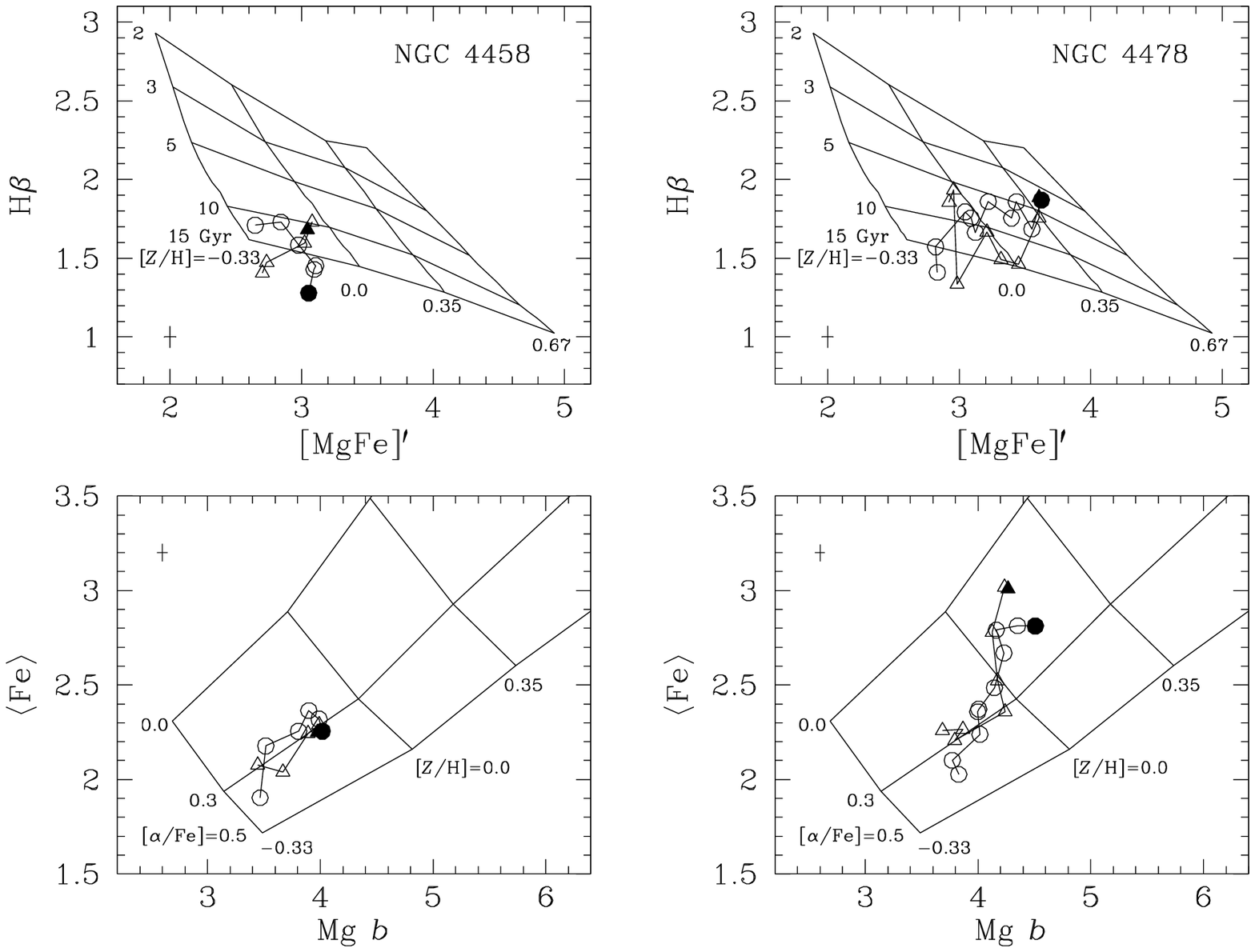,width=\linewidth}
\end{minipage}
\caption{{\it Top panels\/}: The line strength of the \Hb\ and \MgFe\ indices
  at different positions in the galaxies NGC 4458 ({\it left panel\/})
  and NGC 4478 ({\it right panel\/}).  The grid shows the models of
  Thomas et al. (2003) for different metallicities [\ZH] and
  ages. {\it Bottom panels\/}: The line strength of the \Mgb\ and \Fe\
  indices at different positions in the galaxies NGC 4458 ({\it left
  panel\/}) and NGC 4478 ({\it right panel\/}).  In all panels the
  {\it circles\/} show the major axis, the {\it triangles\/} the minor
  axis, the {\it filled symbols\/} the galaxy centres.  The grid shows
  the models of Thomas et al. (2003) for different overabundances
  [\aFe] and metallicities [\ZH], with age 12 Gyr.
\label{fig:mgbfe}}
\end{figure*}

We show in the top panels of Figure \ref{fig:mgbfe} the line strengths of
the \Hb\ and \MgFe\ indices at different positions in the galaxies NGC
4458 and NGC 4478 with the grid of models of TMB. No clear trend is
observed for NGC 4458.  NGC 4478 shows a negative gradient in both the
indices \Hb\ and \MgFe .  We show in bottom panels of Figure
\ref{fig:mgbfe} the line strength of the \Mgb\ and
\Fe\ indices at different positions in the galaxies NGC 4458 and NGC
4478.  For both galaxies, both indices \Fe\ and \Mgb\ decrease
outward on the mean.

\begin{figure}
    \epsfig{file=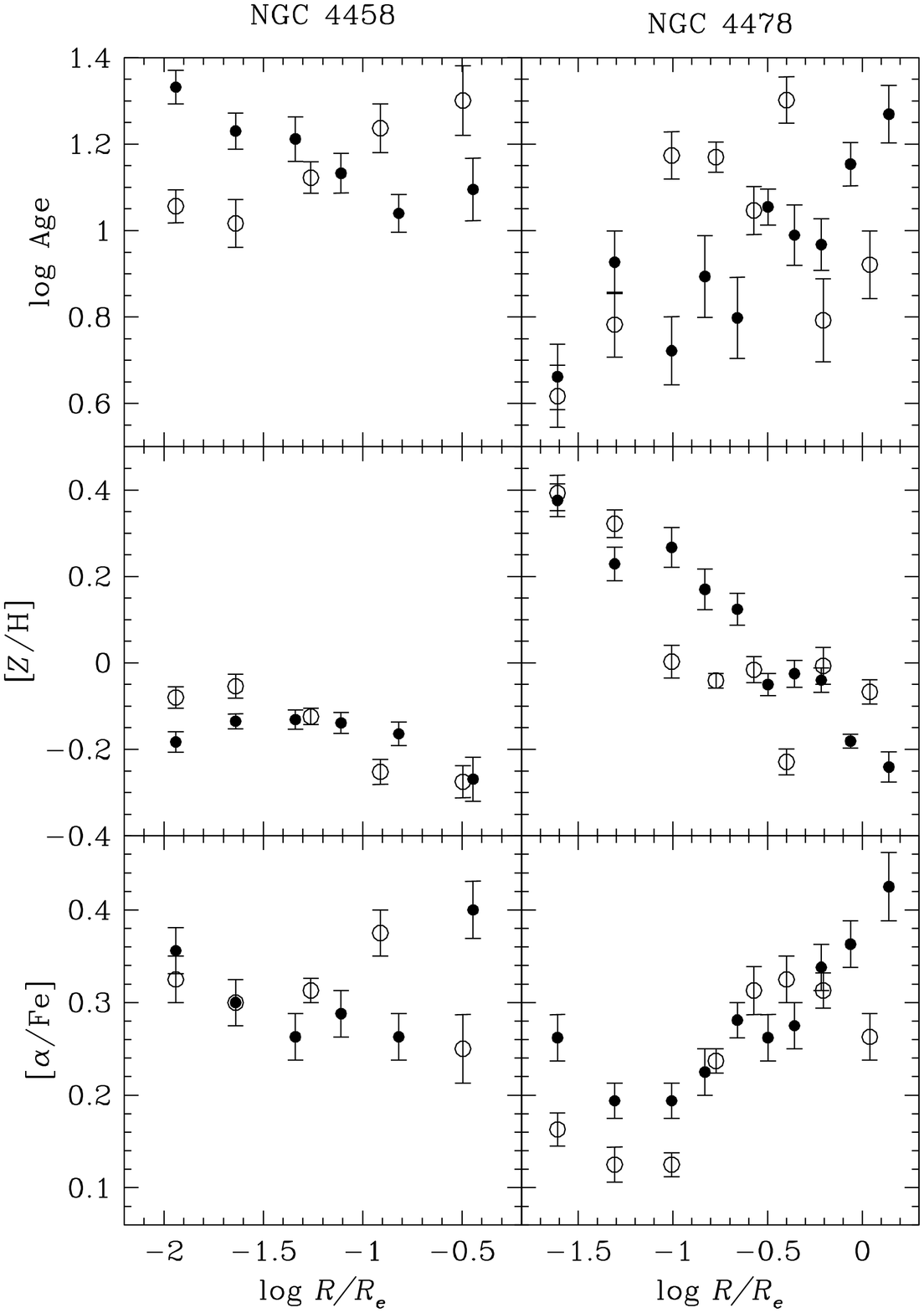,width=\linewidth}
\caption{The age ({\it top panels\/}), metallicity [\ZH] ({\it middle panels\/}),
  and overabundance [\aFe] profiles ({\it bottom panels\/}) as a
  function of the distance from the centre in units of half-luminosity
  radius derived from the \Hb, \Fe\ and \Mgb\ values using the models
  of Thomas et al. (2003).  The {\it filled symbols\/} refer to the
  major axis of the galaxies, the {\it open symbols\/} to the minor
  axis.
\label{fig:agemetover}}
\end{figure}

Using stellar population models with variable element abundance ratios
from TMB we derive average ages, metallicities and [\aFe]\ ratios as a
function of the position in the galaxies, following the procedure of
Mehlert et al. (2003). Figure \ref{fig:agemetover} translates the
trends observed in Figure \ref{fig:mgbfe} as follows. NGC 4458 is
uniformly old ($t\approx 15$ Gyr), has overall low metal content
([\ZH]~$\approx -0.2$) and overabundance ([\aFe]~$\approx 0.3$). For
NGC 4478 instead the central regions ($R<0.1R_e$) are younger
(t~$\approx 6$ Gyr), more metal rich ([\ZH]~$\approx 0.35$) and less
overabundant ([\aFe]~$\approx 0.2$) than the outer regions, where old
ages (t~$\approx 15$ Gyr), low metallicities ([\ZH]~$\approx 0.1$) and
high overabundances ([\aFe]~$\approx 0.3$) are observed on both the
minor and the major axis. {  The size of the error bars shows that
the decreases in age and overabundance in the central region are
significant. Finally, note that within the errors most of the ages
shown in Fig. \ref{fig:agemetover} are compatible with the age of
Universe favoured by WMAP (Spergel et al. 2003).}
In the following we will interpret the radial variations described
above as the result of the superposition of the central discs detected
photometrically in Section \ref{sec:detection} and the main body of
the galaxy.  {  Moreover, we will assume that the main body of the
galaxy does not have a [\aFe ] gradient, in agreement with the results
of Mehlert et al. (2003). }
NGC 4458 has properties similar to giant ellipticals with decoupled
cores. Prototypes of decoupled cores in giant ellipticals are as old
as and metal richer than the rest of the galaxy, with high and
approximately constant overabundance, as in the case of NGC 4816 and
IC 4051 (Mehlert et al. 1998), or NGC 4365 (Surma \& Bender 1995;
Davies et al. 2001).  In NGC 4458 the inner disc and the main body of
the galaxy appear to have the same stellar populations, with a { 
metallicity gradient for the main body of the galaxy slightly weaker
than the cases discussed above.}
By contrast, the case of NGC 4478 is intriguing. The inner, cold disc
is younger, richer in metals and less overabundant than the main body
of the galaxy. We tested this conclusion with a composite stellar
population model, summing 10\% of a young ($t=1.65$ Gyr) and metal rich
($Z=3 Z_\odot$) component with solar \aFe\ ratios, and 90\% of an old
($t=12$ Gyr) metal rich ($Z=2 Z_\odot$) overabundant ([\aFe ]~$=0.4$)
population, and recovering the observed central values of the line
indices. Given the errors of the available spectroscopic data and
their spatial resolution, we refrained from a detailed study of the
allowed range of alternative solutions, that only spectroscopy at HST
resolution can pin down. 

{  Of course, higher than solar \aFe\ ratios for the disk are
allowed if we relax the assumption that the \aFe\ ratio of the main
body of NGC 4478 is constant.}
Using these age and metallicity estimates and the models of Maraston
(1998), we compute that the mass-to-light ratios of the inner discs
are $M/L_V=7.2$ M$_\odot/$L$_\odot$ for NGC 4458 (with a range from 5
to 10 judged from the range of ages and metallicities shows in
Fig. \ref{fig:agemetover}) and $M/L_V=1.3$ M$_\odot/$L$_\odot$ for NGC
4478. In this case, given the limitations of the spectroscopic data
discussed above, the uncertainties are probably as large as a factor
2.  Combined with the luminosities given in Table
\ref{tab:disk_parameters}, this gives masses of $M=3.4\times10^7$
M$_\odot$ and $M=2.5\times10^7$ M$_\odot$, respectively.
The kinematic and photometric properties discussed above argue for the
presence of (cold) discs at the centres of NGC 4458 and NGC 4478, and
therefore for a formation scenario through the dissipational collapse
of a gas rich object. The observed kinematic signature of
counter-rotation in NGC 4458 argues in addition for an external origin
with decoupled angular momentum.  The absence of gradients in the
stellar population of NGC 4458 indicates that the formation of the
inner cold structure happened at the same time as the main body of the
galaxy. The younger age and low overabundance of the central structure
of NGC 4478 is indicative of a prolonged star formation history (Thomas,
Greggio \& Bender 1999), typical of an undisturbed disc-like, gas-rich
(possibly pre-enriched) structure. This is consistent with the absence
of counter-rotation.

\section{Conclusions}
\label{sec:conclusions}

We have investigated the properties of the central regions of the
low-luminosity ellipticals NGC 4458 and NGC 4478, finding that:

\begin{enumerate}

\item HST archive images of the two galaxies show 
the presence of central discs.  They have scale lengths and face-on
central surface brightnesses which fit on $\mu_0^c$--$h$ relation for
galaxy discs. For NGC 4458 these parameters are typical of NSDs, while
for NGC 4478 they are intermediate between the values for NSDs and the
discs of discy ellipticals.

\item Measurements presented by Halliday et al. (2001) 
along the major and minor axis of the galaxies revealed the presence of
decoupled kinematics on the radial scales of the NSDs. A
counter-rotating central structure is present in NGC 4458, a cold
rotating component in NGC 4478.

\item The Lick/IDS absorption line-strength 
indices \Hb, \Mgb\ and \Fe\ measured as a function of radius along the
major and the minor axis of the galaxies allow us to estimate the age,
metallicity and overabundance of the stellar populations of the two
galaxies using simple stellar population models.

\item The radial variations of the derived quantities constrain 
the stellar population properties of the central discs. The NSD of NGC
4458 has an estimated stellar mass of $3.4\times 10^7$ M$_\odot$ and
is similar to the decoupled cores of bright ellipticals, being as old
as and richer in metals than the rest of the galaxy, with high and
approximately constant overabundance.  The cold central disc of NGC
4478 has a similar estimated stellar mass of $2.5\times 10^7$
M$_\odot$, but is younger, richer in metals and less overabundant than
the main body of the galaxy.

\item The nearly solar {  $\alpha-$element abundance} 
of the central disc of NGC 4478 indicates a prolonged star formation
history, typical of an undisturbed disc-like, gas-rich (possibly
pre-enriched) structure.

\end{enumerate}

\section*{Acknowledgments}

\noindent
Based on observations made with the NASA/ESA Hubble Space Telescope,
obtained from the data archive at the Space Telescope Institute. STScI
is operated by the association of Universities for Research in
Astronomy, Inc. under the NASA contract NAS 5-26555.
We thank the observing committee of the CfA for their generous award
of telescope time to obtain the spectroscopic observations presented
here. The spectroscopic absorption line-strength measurements
presented were derived as part of the Ph.D. thesis research of CH. 
CH gratefully acknowledges the support of a PPARC studentship at the
University of Durham and the grant support of the Italian National
Research Centre (grant CNRG00887) and a grant from the Fondo per gli
Investimenti della Ricerca di Base of the Italian Ministery of
Education, University and Research (grant RBAU018Y7E). 
{  We acknowledge A. W. Graham for useful discussion.}
LM gratefully acknowledges the support of an ESO studentship at the
ESO Research Facilities in Santiago.
RPS thanks the financial support by the DFG grant SFB375.
This research has made use of the Lyon-Meudon Extragalactic Database
(LEDA), NASA/IPAC Extragalactic Database (NED) and Starlink facilities.

\appendix


\begin{thebibliography}{}

\bibitem[Arnold et al. 1994]{1994MNRAS.271..924A} Arnold, R., de Zeeuw, P. T., Hunter, C., 1994, \mnras, 271, 984A

\bibitem[Bender et al.(1989)]{1989A&A...217...35B} Bender, R., Surma, P.,
  Doebereiner, S., Moellenhoff, C., \& Madejsky, R., 1989, \aap, 217,
  35

\bibitem[Bertola \& Corsini(1999)]{1999IAUS..186..149B} Bertola, F.~\& 
  Corsini, E.~M.\ 1999, IAU Symp.~186: Galaxy Interactions at Low and
  High Redshift, 186, 149

\bibitem[Bertola et al.(1999)]{1999ApJ...519L.127B} Bertola, F., Corsini,
  E.~M., Vega Beltr{\' a}n, J.~C., Pizzella, A., Sarzi, M.,
  Cappellari, M., \& Funes, J.~G., 1999, \apjl, 519, L127

\bibitem[Balcells, Graham, Dom{\'{\i}}nguez-Palmero, \&
  Peletier(2003)]{2003ApJ...582L..79B} Balcells, M., Graham, A.~W.,
  Dom{\'{\i}}nguez-Palmero, L., \& Peletier, R.~F., 2003, \apjl, 582,
  L79

\bibitem[Balcells, Graham, \&
  Peletier(2004)]{bal2004} Balcells, M., Graham, A.~W.,
  \& Peletier, R.~F., 2004, ApJ, submitted (astro-ph/0404381)

\bibitem[B{\" o}ker et al.(2002)]{2002AJ....123.1389B} B{\" o}ker, T.,
  Laine, S., van der Marel, R.~P., Sarzi, M., Rix, H., Ho, L.~C., \&
  Shields, J.~C.\ 2002, \aj, 123, 1389

\bibitem[Bruzual \& Charlot(2003)]{2003MNRAS.344.1000B} Bruzual, G.~\& 
  Charlot, S., 2003, \mnras, 344, 1000

\bibitem[Cappellari(2002)]{cap02} Cappellari, M.\ 2002,
  \mnras, 333, 400

\bibitem[Caon, Capaccioli, \& Rampazzo(1990)]{1990A&AS...86..429C} Caon,
  N., Capaccioli, M., \& Rampazzo, R., 1990, \aaps, 86, 429

\bibitem[Carollo, Danziger, Rich, \& Chen(1997)]{Carollo97} Carollo,
  C.~M., Danziger, I.~J., Rich, R.~M., \& Chen, X., 1997, \apj, 491,
  545

\bibitem[Davies et~al., 1993]{dav93}
  Davies, R.~L., Sadler, E.~M., and Peletier, R.~F., 1993, \mnras, 262, 650

\bibitem[Davies et al. (2001)]{detal01}
  Davies, R.L., Kuntschner, H., Emsellem, E., Bacon , R., Bureau, M.,
  Carollo, C.M., Copin, Y., Miller, B.W., Monnet, G., Peletier, R.F.,
  Verolme, E.K., de Zeeuw, P.T., 2001, ApJL, 548, L33

\bibitem[de Vaucouleurs et al.(1991)]{1991RC3...C......0D}
  de Vaucouleurs, G., de Vaucouleurs, A., Corwin, J.~R., Buta, R.~J.,
  Paturel, G., \& Fouque, P.\ 1991, Third Reference Catalogue of
  Bright Galaxies (RC3)

\bibitem[Faber et~al., 1985]{fab85}
  Faber, S.~M., Friel, E.~D., Burstein, D., and Gaskell, C.~M., 1985,
  \apjs, 57, 711

\bibitem[Faber et al.(1997)]{1997AJ....114.1771F} Faber, S. M.,
  Tremaine, Scott, Ajhar, Edward A., Byun, Yong-Ik, Dressler, Alan,
  Gebhardt, Karl, Grillmair, Carl, Kormendy, John, Lauer, Tod R.,
  Richstone, Douglas, 1997, \aj, 114, 1771

\bibitem[Ferrarese et al.]{ferra1994} 
  Ferrarese, L., van den Bosch, F. C., Ford, H. C., Jaffe, W.,
  O'Connell, R. W., 1994, \aj, 108, 1598F

\bibitem[Gonz\'{a}lez, J.~J. (1993)]{gon93} Gonz\'{a}lez, J.~J.
  1993, Line Strength Gradients and Kinematic Profiles in Elliptical
  Galaxies, PhD thesis, University of California

\bibitem[Gorgas et~al. (1990)]{gor90} Gorgas, J., Efstathiou, G.,
  \& Arag\'{o}n-Salamanca, A.~A., 1990, MNRAS, 245, 217

\bibitem[Graham(2001)]{2001MNRAS.326..543G} Graham, A.~W.\ 2001, 
  \mnras, 326, 543

\bibitem[Graham \& Guzm{\' a}n(2003)]{2003AJ....125.2936G} 
  Graham, A.~W.~\& Guzm{\' a}n, R.\ 2003, \aj, 125, 2936

\bibitem[Halliday(1999)]{hal99} Halliday, C. 1999,
  Low-Luminosity Elliptical Galaxies, PhD thesis, University of
  Durham

\bibitem[Halliday et~al. (2001)]{hal01} Halliday, C., Davies,
  R.~L., Kuntschner, H., Bender, R., Birkinshaw,
  M., Saglia, R.~P., \& Baggley, G. 2001, MNRAS, 326, 473

\bibitem[Haynes et al. (1999)]{1997AJ....117.2039F} Haynes, M.P.,
  Giovanelli, R., Chamaraux, P., da Costa, L.N., Freudling, W.,
  Salzer, J.J., Wegner, G., 1997, AJ, 117, 2039

\bibitem[Holley-Bockelmann \& Richstone(2000)]{2000ApJ...531..232H}
Holley-Bockelmann, K.~\& Richstone, D.~O., 2000, \apj, 531, 232

\bibitem[Holtzman et al.(1995)]{hol95a} Holtzman, J.~A.~et
  al.\ 1995a, PASP, 107, 156

\bibitem[Holtzman et al.(1995)]{hol95b} Holtzman, J.~A.,
  Burrows, C.~J., Casertano, S., Hester, J.~J., Trauger, J.~T.,
  Watson, A.~M., \& Worthey, G.\ 1995b, PASP, 107, 1065

\bibitem[Jaffe et al.]{jaffe94} 
  Jaffe, W., Ford, H. C., O'Connell, R. W., van den Bosch, F. C.,
  Ferrarese, L., 1994, \aj, 108, 1567J

\bibitem[Jedrzejewski]{Jedrz87} Jedrzejewski, R., 1987, \mnras, 226, 747J

\bibitem[Kinney et al.(1996)]{1996ApJ...467...38K} Kinney, A.~L., Calzetti,
  D., Bohlin, R.~C., McQuade, K., Storchi-Bergmann, T., \& Schmitt,
  H.~R.\ 1996, ApJ, 467, 38

\bibitem[Kobayashi \& Arimoto(1999)]{1999ApJ...527..573K} 
  Kobayashi, C.~\& Arimoto, N., 1999, \apj, 527, 573

\bibitem[Kormendy, 1984]{kor84} Kormendy, J., 1984,\apj, 287, 577

\bibitem[Kormendy et al.(1996)]{1996IAUS..171..105K} Kormendy, J.,
  Byun, Y.; Ajhar, E. A.; Lauer, T. R.; Dressler, A.; Faber, S. M.;
  Grillmair, C.; Gebhardt, K.; Richstone, D.; Tremaine, S., 1996, IAU
  Symp.~171: New Light on Galaxy Evolution, 171, 105

\bibitem[Kormendy \& Bender(1996)]{1996ApJ...464L.119K} Kormendy, J.~\&
  Bender, R., 1996, \apjl, 464, L119

\bibitem[Kormendy \& Gebhardt(2001)]{2001AIPC..586..363K} Kormendy, J.~\&
  Gebhardt, K., 2001, AIP Conf.~Proc.~586: 20th Texas Symposium on
  Relativistic Astrophysics, 586, 363

\bibitem[Krist \& Hook (1999)]{kri99} Krist, J., Hook, R. 1999,
  STIS Instrument Handbook Version 4.0 (Baltimore: STScI)

\bibitem[Lauer et al.(1995)]{1995AJ....110.2622L} Lauer, T. R., Ajhar,
  E. A., Byun, Y.-I., Dressler, A., Faber, S. M., Grillmair, C.,
  Kormendy, J., Richstone, D., Tremaine, S., 1995, \aj, 110, 2622

\bibitem[1998]{Mar98} Maraston, C., 1998, MNRAS, 300, 872

\bibitem[1998]{Metal98} Mehlert, D., Saglia, R.P., Bender, R., Wegner, G., 1998, 
  A\&A, 332, 33

\bibitem[Mehlert, Saglia, Bender, \& Wegner(2000)]{2000A&AS..141..449M} 
  Mehlert, D., Saglia, R.~P., Bender, R., \& Wegner, G., 2000, \aaps,
  141, 449

\bibitem[2003]{Metal03}
  Mehlert, D., Thomas, D., Saglia, R.P., Bender, R., Wegner, G., 2003,
  A\&A, 407, 423

\bibitem[Michard(1985)]{1985A&AS...59..205M} Michard, R., 1985, \aaps, 59,
  205

\bibitem[Monnet et al.(1992)]{mon92}
  Monnet, G., Bacon, R. \& Emsellem, E.\ 1992, A\&A, 253, 366

\bibitem[Peletier et al. (1989)]{pel89} Peletier, R.~F. 1989,
  PhD thesis, University of Groningen

\bibitem[Peletier et al.(1990)]{1990AJ....100.1091P} Peletier, R.~F.,
  Davies, R.~L., Illingworth, G.~D., Davis, L.~E., \& Cawson, M.,
  1990, \aj, 100, 1091

\bibitem[Pizzella et al.(2002)]{piz02} Pizzella, A., Corsini, E. M.,
  Morelli, L., Sarzi, M., Scarlata, C., Stiavelli, M.,\& Bertola,
  F. 2002, ApJ, 573, 131

\bibitem[Press et al.(1992)]{pres92} Press, W.~H., Teukolsky, S.~A.,
  Vetterling, W.~T., \& Flannery, B.~P.\ 1992, Numerical recipes in
  FORTRAN. The art of scientific computing (Cambridge: University
  Press)

\bibitem[Prugniel, Nieto, \& Simien(1987)]{1987A&A...173...49P} Prugniel,
  P., Nieto, J.-L., \& Simien, F., 1987, \aap, 173, 49

\bibitem[Ravindranath et al.(2001)]{2001AJ....122.653R}
  Ravindranath, S., Ho, L. C., Peng, C. Y., Filippenko, A. V., Sargent,
  W. L. W., 2001, \aj, 122, 653

\bibitem[Rest et al.(2001)]{2001AJ....121.2431R} Rest, A., van den
  Bosch, F.~C., Jaffe, W., Tran, H., Tsvetanov, Z., Ford, H.~C.,
  Davies, J., \& Schafer, J., 2001, \aj, 121, 2431

\bibitem[Rix and White, 1992]{rix92} Rix, H.~W. and White, S. D.~M., 1992, 
  \mnras, 254, 389

\bibitem[Sandage \& Tammann(1981)]{1981rsac.book.....S} Sandage, A.~\&
  Tammann, G.~A.\ 1981, Revised Shapley-Ames Catalog of Bright
  Galaxies, (Washington: Carnegie Institution) (RSA)

\bibitem[Scorza \& Bender (1995)]{sco95}
  Scorza, C.\ \& Bender, R.\ 1995, A\&A, 293, 20 (SB95)

\bibitem[Scorza \& van den Bosch (1998)]{sco98}
  Scorza, C.\ \& van den Bosch, F.\ C.\ 1998, MNRAS, 300, 469

\bibitem[Seifert \& Scorza (1996)]{sei96}
  Seifert, W.\ \& Scorza, C.\ 1996, A\&A, 310, 75 (SS96)

\bibitem[Spergel et al.(2003)]{spe03} 
  Spergel, D. N., Verde, L., Peiris, H. V., Komatsu, E., Nolta, M. R.,
  Bennett, C. L., Halpern, M., Hinshaw, G., Jarosik, N., Kogut, A.,
  Limon, M., Meyer, S. S., Page, L., Tucker, G. S., Weiland, J. L.,
  Wollack, E., \& Wright, E. L. 2003, ApJS, 148, 175

\bibitem[Surma \& Bender (1995)]{sub95}
  Surma, P.\ \& Bender, R.\ 1995, A\&A, 298, 405

\bibitem[Tantalo, Chiosi, \& Bressan(1998)]{1998A&A...333..419T} 
  Tantalo, R., Chiosi, C., \& Bressan, A., 1998, \aap, 333, 419

\bibitem[1999]{TGB99}
  Thomas, D., Greggio, L, Bender, R.,  1999 MNRAS, 302, 537

\bibitem[2003]{Tetal03a}
  Thomas D., Maraston, C., Bender R. 2003, MNRAS, 339, 897 (TMB)

\bibitem[Tomita et al.(2000)]{tomi00} Tomita, A., Aoki, K., 
  Watanabe, M., Takata, T., Ichikawa, S., 2000, \aj, 120, 123

\bibitem[Trager (1997)]{tra97} Trager, S.~C. 1997
  The Stellar Population Histories of Elliptical Galaxies,
  PhD thesis, University of California

\bibitem[Trager et al. (1998)]{tra98} Trager, S.~C., Worthey, G.,
  Faber, S.~M., Burstein, D., \& Gonz\'{ a}lez, J.~J.\ 1998, ApJS,
  116, 1

\bibitem[Tran et al.(2001)]{tran01} Tran, H. D., 
  Tsvetanov, Z., Ford, H. C., Davies, J., Jaffe, W., van den Bosch,
  F. C., Rest, A., 2001, \aj, 121, 2928

\bibitem[Trujillo, Erwin, Ramos, \& Graham(2004)]{2004AJ....127.1917T} 
  Trujillo, I., Erwin, P., Ramos, A.~A., \& Graham, A.~W.\ 2004, \aj,
  127, 1917

\bibitem[Tully (1988)]{1988ngc..book.....T}
  Tully, R.\ B.\ 1988, Nearby Galaxy Catalog, (Cambridge: Cambridge
  University Press)

\bibitem[van den Bosch et al. (1994)]{vdb94}
  van den Bosch, F.~C., Ferrarese, L., Jaffe, W. Ford, H.C.,
  O'Connell, R.W., 1994, \aj, 108, 1579

\bibitem[van den Bosch(1998)]{vdb98a}
  van den Bosch, F.~C.\ 1998, ApJ, 507, 601

\bibitem[van den Bosch \& Emsellem(1998)]{1998MNRAS.298..267V} van den
  Bosch, F.~C.~\& Emsellem, E., 1998, \mnras, 298, 267

\bibitem[van den Bosch et al.(1998)]{vdb98b} van den Bosch, F.\ C.,
  Jaffe, W. \& van der Marel, R.\ P.\ 1998, \mnras, 293, 343

\bibitem[van Dokkum, P. G.\& Franx, M.]{vadok98} van Dokkum, P. G.\& Franx, M.,
  1995, \aj, 110, 2027V

\bibitem[Vazdekis(1999)]{1999ApJ...513..224V} Vazdekis, A., 1999, \apj, 
  513, 224

\bibitem[Worthey, 1994]{wor94} Worthey, G., 1994, \apjs, 95,107

\end{thebibliography}
\end{document}